# Carbon ablators with porosity designed for enhanced aerospace thermal protection

Erik Poloni[1], Florian Bouville[1,+], Alexander L. Schmid[1], Pedro I. B. G. B. Pelissari[2], Victor C. Pandolfelli[2], Marcelo L. C. Sousa[1], Elena Tervoort[1], George Christidis[3], Valery Shklover[3], Juerg Leuthold[3], André R. Studart[1,*]

[1] Complex Materials, Department of Materials, ETH Zürich, 8093 Zürich, Switzerland

[2] Federal University of São Carlos, Materials Engineering Department, São Carlos, SP, Brazil

[3] Institute of Electromagnetic Fields, ETH Zürich, 8092 Zürich, Switzerland

[+] Now at: Centre for Advanced Structural Ceramics, Department of Materials, Imperial College London, United Kingdom

[*] Corresponding author. Tel: +41 44 633 70 50. E-mail: andre.studart@mat.ethz.ch.

1. Abstract

Porous carbon ablators offer cost-effective thermal protection for aerospace vehicles during re-entry into planetary atmospheres. However, the exploration of more distant planets requires the development of ablators that are able to withstand stronger thermal radiation conditions. Here, we report the development of bio-inspired porous carbon insulators with pore sizes that are deliberately tuned to enhance heat-shielding performance by increasing scattering of high-temperature thermal radiation. Pore size intervals that promote scattering are first estimated using an established model for the radiative contribution to the thermal conductivity of porous insulators. On the basis of this theoretical analysis, we identify a polymer additive that enables the formation of pores in the desired size range through the polymerization-induced phase separation of a mixture of phenolic resin and ethylene glycol. Optical and electron microscopy, porosimetry and mechanical tests are used to characterize the structure and properties of porous insulators prepared with different resin formulations. Insulators with pore sizes in the optimal scattering range reduce laser-induced damage of the porous structures by up to 42%, thus offering a promising and simple route for the fabrication of carbon ablators for enhanced thermal protection at high temperatures.

## 2. Introduction

The increasing interest in space exploration in the public and private sectors has motivated the development of new material systems that can withstand the extreme conditions of hypersonic flights. [1, 2] Ceramic matrix composites, [3] ultra-high temperature ceramics [4] and carbon-based ablators [5] are some of the materials used for thermal protection in aerospace applications. Thermal protection systems are crucial to ensure safe entry of space vehicles and probes in the atmosphere of planets and to shield vehicles from the heat generated by propulsion systems. [2, 6] During entry into the atmosphere, friction between the outer gas and the surface of the hypersonic vehicle causes extreme local heating and the formation of a boundary layer of ionized gas at the leading edge. [1] To cope with these extreme conditions, polymer-based ablative materials that consume part of the heat upon pyrolysis are often used as thermal protection system. [5] Phenolic impregnated carbon ablators (PICA) are an example of a thermal protection system that has been successfully employed for lunar and Martian missions. [7, 8] In addition to a weight reduction of 50% compared to previous heat shields, PICA also relies on a precursor infiltration approach that is simple, cost-effective and scalable. However, future planetary missions will impose even more challenging requirements for thermal protection systems, since heat transfer will likely be dominated by radiation instead of convection. [9] The thin and $CO_2$-rich atmosphere of Mars and the ever larger spacecrafts, for example, should lead to a much stronger radiation during re-entry, which cannot be effectively addressed by state-of-the-art heat shield materials.

Several approaches have been proposed for the design of thermal protection systems showing low thermal conductivity and reduced radiative heat transfer in the near infrared wavelength range. These include reflective woven fibres with photonic bandgaps, [10, 11] reflective flakes, [12] porous materials, [13, 14] periodic woodpile structures and reflective porous glasses. [15] High-temperature photonic structures have also been developed to tune the optical properties of thermal barrier coatings for turbine applications [16, 17] or emitters for thermophotovoltaic technologies. [18, 19] Such materials can potentially also be used for thermal radiation protection if incorporated in the form of photonic filler particles into conventional ablators.

Following this rationale, multilayered metal-dielectric structures have been designed and fabricated as a possible route to create photonic additives for enhanced radiative thermal protection. [12] Recent work has also shown the possibility to manufacture polymer fibers with a highly reflective titania coating, [20] which could in principle also be applied for the fabrication of fiber-based composites with improved near infrared reflectance. Although these reflective and photonic bandgap structures may offer prospective solutions for the next generation of heat shields, their implementation in thermal protection systems remains challenging due to the high cost, cumbersome fabrication process, insufficient high temperature resistance or limited scalability of the technologies developed so far.

Heat shields for more demanding planetary missions may also be developed by enhancing the reflective and scattering properties of state-of-the-art carbon-based ablators, thus taking advantage of the high temperature resistance of carbon and of the scalability, low cost and readiness of current manufacturing technologies. State-of-the-art PICA is manufactured by impregnating a phenolic resin into a porous network of carbon fibers, [21] followed by thermal polymerization and removal of the processing solvents. [7] Because most research and development on carbon ablators have been driven by space agencies and companies, very limited information is available in the literature on the processing-structure-property relationships of these materials. [22] Microstructural analysis of as-fabricated PICA and related formulations indicates that the phenolic resin forms an aerogel-like porous structure between the carbon fibers upon thermal curing (Figure 1a). [23, 24] Carbon ablators with an aerogel-filled fibrous structure have recently been shown to effectively withstand the harsh thermal conditions in arc jet tests that simulate atmospheric re-entry conditions. [22, 25] Porosity in these materials is generated via a self-assembly process that relies on the polymerization-induced phase separation of the resin during curing. While the aerogel structure is clearly important for the thermal insulation properties of such fibrous materials, [24] a systematic evaluation of the role of porosity and how it can be possibly tuned to reduce radiative heat transfer in carbon ablators has not yet been reported. In this context, living organisms provide inspiring examples of how reflective photonic structures can be created using up-scalable self-assembly processes. [26-30] For example, *Cyphochilus* beetles produce scales with a photonic chitinous network that has been evolutionary optimized for the multiple scattering of white light. The development of processing routes that enable the incorporation of such structures in synthetic materials may lead to thermal protection systems with enhanced scattering efficiency and radiative heat transfer resistance.

Here, we develop and study an up-scalable self-assembly route to control the porous architecture and enhance the radiative thermal resistance of a phenolic impregnated carbon ablator. Porosity is introduced in conventional PICA using a polymer that directly influences the phase separation process that takes place during thermal treatment of the phenolic resin. To identify the pore sizes required to promote radiation scattering in the near infrared range, we first perform a theoretical analysis of the thermal properties expected for the porous structures at high temperatures. Phase separation of the resin in the presence of different fractions of polymer and solvent is then thoroughly investigated to enable the manufacturing of carbon ablators with the desired average pore size. Finally, we measure the mechanical properties and thermal radiation resistance of the fibrous structures to quantify the effect of the controlled porosity on the performance of the carbon ablator.

## 3. Results and Discussion
### 3.1. *Theoretical analysis of radiative transfer*

State-of-the-art carbon ablators rely on the aerogel formed between the carbon fiber network to reduce radiative transfer through the structure at high temperatures (Figure 1a). [7] While the formation of the aerogel has been shown to enhance the ablation resistance of the structure, the reported sizes of the pores in the aerogel are below 100 nm [22, 24, 25] and thus much smaller than the micrometer wavelengths ($\lambda$) of the incoming and re-emitted radiation at high-temperature entries. For temperatures ranging from 1000 to 3500°C, the incoming and re-emission radiation is expected to peak at wavelengths between 0.77 and 2.28 µm, as predicted by Wien's law for black-body radiation. [31, 32] This suggests that the pore size of state-of-the-art ablators reported in the literature might not be in a suitable range to promote strong scattering of the thermal radiation. Indeed, strongest radiation scattering is expected to take place when the pore size is comparable to the wavelength corresponding to maximum black-body radiation. [13] To test this hypothesis, we theoretically estimate the effect of the pore size on the thermal conductivity of carbon ablators at temperatures in the range 1000–3500°C, following analytical models outlined in a previous study. [32] It is important to note that the model is used to predict the thermal conductivity of the material and not its behavior in the specific conditions of planetary entry or of the laser experiment performed later in our study. Therefore, the theoretical analysis considers scattering in all directions of the radiation generated by the material at high temperatures. Our basic assumption is that the ablation resistance should be enhanced if the material shows low thermal conductivity at the high temperatures expected during the laser test and planetary entry conditions. This keeps heat at the surface of the ablator, effectively shielding the vehicle from thermal damage.

The thermal conductivity of porous materials at these high temperatures is dominated by thermal radiation and therefore depends strongly on the size and density of light scattering pores present in the structure. [32] The radiative contribution to the thermal conductivity can be expressed by the following relation:

$$k_{rad} = \frac{16\sigma n^2 T^3}{3\beta_R} \qquad (Eq.\ 1)$$

where $\sigma$ is the Stefan-Boltzmann constant, $n$ is the effective refractive index of the material, $T$ is the absolute temperature and $\beta_R$ is the Rosseland extinction coefficient.

Assuming the material to behave as a black body emitter, one can estimate the Rosseland coefficient from Planck's thermal radiation spectrum and the spectral extinction coefficient of the porous structure, $\beta_{ext}(\lambda)$. [32] To simplify our analysis, we calculate the coefficient $\beta_{ext}$ for the wavelength that corresponds to the maximum thermal radiation emitted by a black body at the considered temperatures. Under this assumption, $\beta_R$ is equal to $\beta_{ext}$ and can be directly estimated from the relation (supporting information):

$$\beta_R = \beta_a + \beta_s \qquad (Eq.\ 2)$$

where $\beta_a$ and $\beta_s$ correspond to the absorption and scattering coefficients, respectively.

Previous experiments indicate that the absorption coefficient ($\beta_a$) of carbon aerogels is four times higher than the scattering coefficient ($\beta_s$) for wavelengths in the range 2–16 $\mu$m and at room temperature. [33] The reduced level of scattering in such aerogels can be explained by the presence of pores much smaller than the typical wavelengths considered in the experiments. However, we expect the scattering contribution to the Rosseland coefficient ($\beta_s$) to play a major role in carbon ablators featuring pore sizes comparable to the wavelength relevant at temperatures above 1000°C ($\approx$ 1 $\mu$m). Therefore, we focus our theoretical analysis on the scattering coefficient ($\beta_s$) and later compare our predictions to the absorption coefficient experimentally measured for porous carbon.

The influence of the porosity and pore size of the carbon ablator on the scattering coefficient, $\beta_s$, can be described using the following relation:

$$\beta_s = Q_s(r) N_p A_p \qquad \text{(Eq. 3)}$$

where $Q_s(r)$ is the scattering efficiency for a pore of radius $r$, $N_p$ is the number of pores per unit volume of material and $A_p$ is the cross-sectional area of the pore. One can show that the product $N_p A_p$ is equal to $3P/(4r)$, with $P$ being the porosity.

Assuming a dominant role of scattering ($\beta_R \approx \beta_s$), this simplified analysis allows us to predict the effect of the pore size ($r$) on the Rosseland extinction coefficient ($\beta_R$) and on the radiative thermal conductivity (Eq. 1) of the porous structure. The dependence of $\beta_R$ on the pore radius is dictated by the product $N_p A_p$ and by the scattering efficiency $Q_s(r)$. For a constant porosity of 0.8, the term $N_p A_p$ scales with $1/r$ and thus decreases for increasing pore sizes. By contrast, numerical calculations based on Mie theory show that $Q_s(r)$ increases monotonically with the pore radius in the range $0.1 < r < 1$ µm, before it levels off at values around 2 for larger pore sizes (Figure 1b). These calculations were carried out for the fixed wavelength $\lambda = 1.4$ µm, which corresponds to the maximum radiation at $T = 1800$ °C, using the open-source algorithm developed by Laven. [34] The effective refractive index $n = 1.24$ of porous carbon was assumed in this analysis, since the resin is totally pyrolyzed into carbon at this high temperature. In another series of simulations (not shown), we observed that our analysis leads to similar conclusions if we use a solid phase with $n = 1.47$, [35] which is the refractive index of the phenolic resin. The rationale behind the choice of these refractive indices is that the heat shielding effect of the ablator arises in a first moment from the virgin porous phenolic structure, and later from the pyrolyzed porous carbon structure as the ablation process proceeds.

The opposing effects of the pore size on $Q_s(r)$ and $N_p A_p$ eventually lead to a maximum in the Rosseland extinction coefficient at $r$ values comparable to the wavelength taken in the analysis. Pores smaller than 100 nm result in a favorable high volumetric density ($N_p$) but are not strong scatterers (low $Q_s$) for the selected wavelength of 1.4 µm. This indicates that the pore sizes of reported aerogels ($r < 100$ nm) are indeed too small to reduce radiative transfer in the ablator at high temperatures. With average pores typically around 70 µm, the non-impregnated carbon fiber network contains highly scattering elements but at an insufficient volumetric number density. Our analysis shows that spherical pore sizes in the range 0.1–10 µm should thus be most effective in reducing the high-temperature thermal conductivity of the ablator. For this pore size range, the scattering coefficient ($\beta_s$) at an arbitrary high temperature of 1800°C varies between 0.25 and 3.7 µm$^{-1}$, which is on the same order of magnitude or up to 10-fold higher than the absorption coefficient previously measured for carbon aerogels for the same temperature (see supporting information). [33] This confirms the relevance of scattering for porous carbon ablators with pore sizes in the range 0.1–10 µm.

To assess the importance of radiative transfer in ablators with pore sizes between 100 nm and 10 µm, we also evaluated the contribution of heat conduction ($k_{cond}$) to the total thermal conductivity ($k_t$) of such optimized porous structures. Assuming convection to play a minor role for these pore sizes, the total thermal conductivity is then given by $k_t = k_{rad} + k_{cond}$. The conductive term $k_{cond}$ was estimated taking into account the relative contributions of the gas and solid phases to the propagation of phonons through the porous structure (see supporting information). [32] Heat conduction in the ablator is expected to depend mainly on the porosity of the structure but not on the pore size (Figure 1c). By comparing the $k_{rad}$ and $k_{cond}$ terms of the total thermal conductivity at 1800°C, we find that pore diameters close to 10 µm generate sufficient scattering to reduce the radiative transfer term to values comparable to those of the conductive contribution. Estimations of the total thermal conductivity of the ablator at other temperatures reveal that the upper limit of the optimum pore size window shifts from approximately 10 µm to 1 µm as the temperature is raised from 1800°C to 3500°C (Figure 1d). According to this analysis, processing routes that enable the formation of pores in the size range 0.1–10 µm have the potential to significantly improve the thermal insulation properties of carbon ablators at temperatures up to 1800°C.

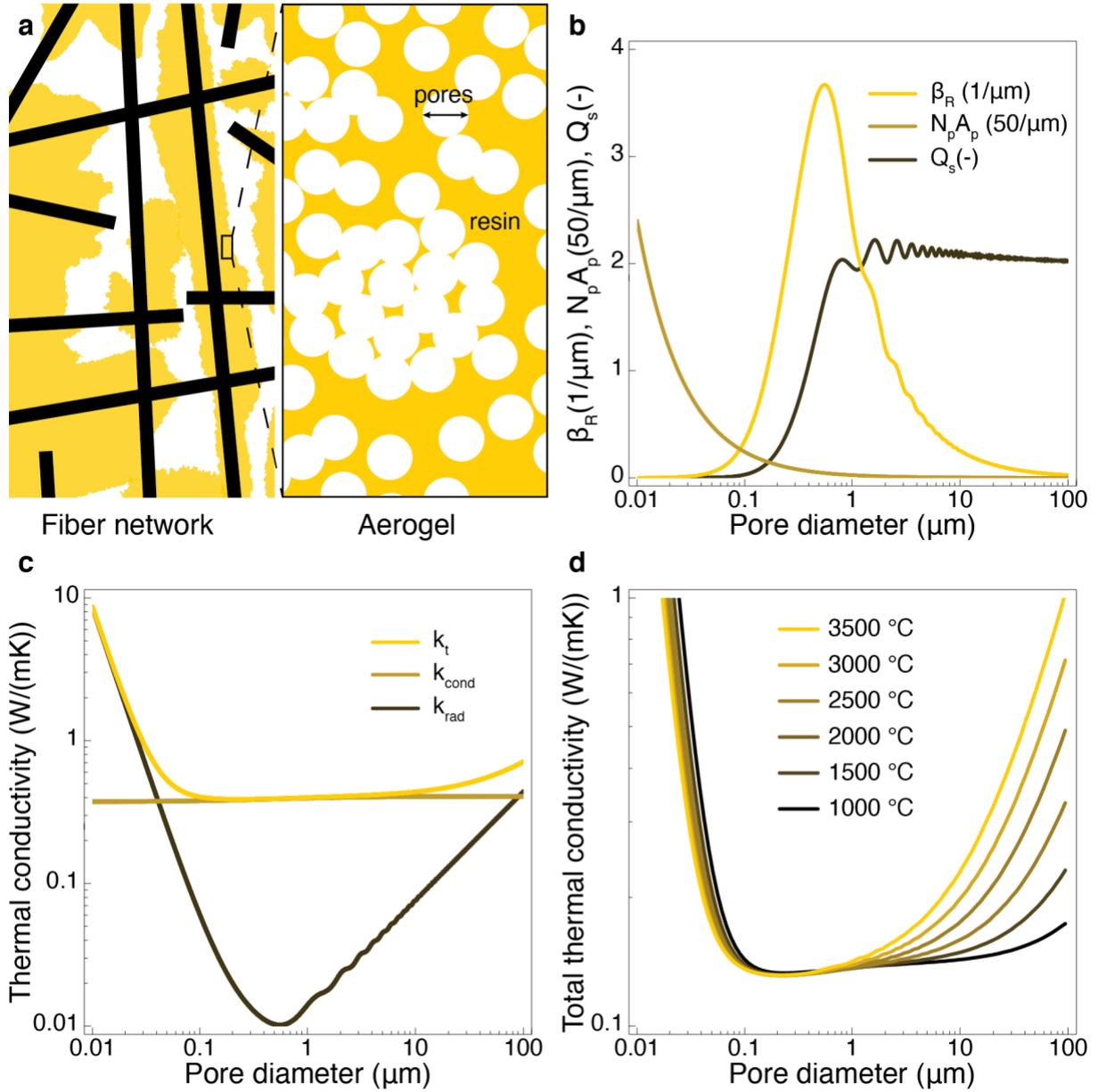

Figure 1: Theoretical predictions for the effect of the pore size on the thermal conductivity of porous carbon ablators. (a) Cartoon showing the network of carbon fibers at the microscale (pores typically ~ 100 μm) and the aerogel-like resin of the porous ablator at the nanoscale (pores typically < 100 nm). (b) Effect of the pore size on the Rosseland extinction coefficient $\beta_R$, which is a product of $Q_s$ and $N_p A_p$, considering a porosity of 0.8. The units are chosen to facilitate the visualization (c) Estimated conduction and radiation contributions to the thermal conductivity of the carbon ablator as a function of the pore diameters. The total thermal conductivity is minimum for pore sizes in the range 0.1–10 μm (d) Influence of the temperature on the total thermal conductivity of carbon ablators with different pore sizes.

## 3.2. Polymerization-induced phase separation

The polymerization-induced phase separation of phenolic resins is a simple and well-established approach to create pores with sizes that typically range from 15 nm to 3 µm. [36-38] To induce spontaneous phase separation upon polymerization, the phenolic resin is first mixed with a good solvent to form a clear homogeneous solution at room temperature (Figure 2). Ethylene glycol is often used as the solvent in such homogeneous mixture. Phase separation occurs when the mixture is heated up to initiate the polymerization of the phenolic resin (Figure 2a). The temperature-triggered chemical reaction of the phenolic precursors leads to macromolecules that are no longer soluble in the solvent. This ultimately results in separation of the initial constituents of the solution into a polymer-rich and a solvent-rich phase (Figure 2b). Chemical additives are often incorporated in the initial solution to influence the phase separation process and thus control the length scale of the polymer- and solvent-rich domains obtained after polymerization. [37] Finally, removal of the solvent by simple evaporation leads to a porous structure with pore sizes comparable to the length scale of the phase-separating domains (Figure 2c).

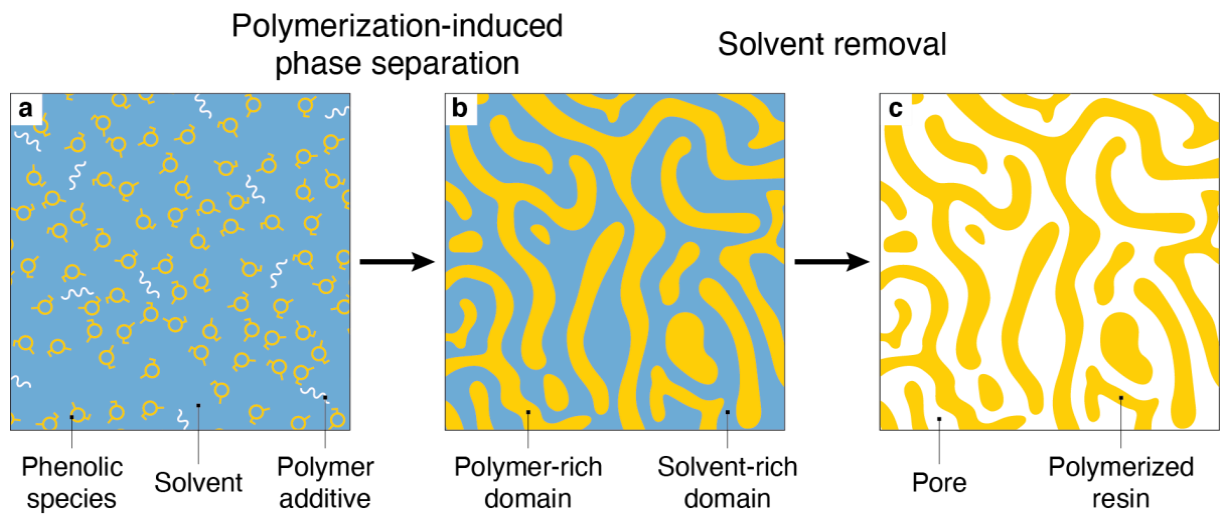

Figure 2: Schematics of the phase separation expected during polymerization of the resin/solvent mixture. (a) Initial mixture of phenolic species and polymer additive dissolved in the solvent. (b) Formation of polymer- and solvent-rich domains upon polymerization of the phenolic species. (c) Pores arising from the solvent-rich domains upon removal of the solvent.

To generate porous structures with pores in the desired size interval of 0.1–10 μm, we experimentally explored resin formulations containing poly(vinylpyrrolidone) (PVP) as polymer additive. PVP was selected as polymer additive because it is known to influence phase separation processes used for the preparation of porous membranes. [39-42] The formulations comprise a base-catalyzed phenolic resin (PR) dissolved in ethylene glycol (EG) at resin:solvent volume ratios varying from 1:2 to 1:7.5. The polymers were pre-dissolved in the initial solution at 110°C in fractions up to 50 wt% relative to the mass of resin. The phase separation process was induced by keeping the resin/solvent/polymer mixture at 150°C for 12 hours. This triggers the polymerization of the resin through condensation reactions between the phenolic molecules.

PVP was found to be effective in promoting the formation of phase-separated domains with length scales on the order of 1 μm or higher. The effect of PVP on the phase separation process was evidenced by the change of the resin/solvent mixture from a transparent solution to a turbid suspension, which happens earlier if the polymer is present in the initial solution. Such a change in the appearance of the mixture indicates the formation of scattering units on the order of the wavelength of visible light. We expect such scattering units to arise from the phase separation of the initial solution into the growing network of polymerized phenolic resin and the solvent phase rich in ethylene glycol.

To better understand the effect of PVP on the polymerization-induced phase separation process, we measured the turbidity change of heated solutions containing different initial fractions of PR, EG and PVP (Figure 3). Because the turbidity is visible with the naked eye, it can be readily quantified from photographs of the solutions over time. The fact that the visible wavelength spectrum falls within the length scale interval that is optimum to reduce thermal radiation (0.1–10 μm) makes this simple analytical approach very convenient for the identification of formulations that might lead to porous ablators with enhanced thermal performance.

Our experiments show that the addition of 10 wt% of PVP to the resin leads to a quick increase in turbidity of solutions prepared with a PR:EG ratio of 1:5 (Figure 3a). We illustrate this effect by displaying the changes in the normalized grayscale value of the image as a function of time for mixtures containing up to 40 wt% PVP and a fixed PR:EG ratio of 1:5 (Figure 3b). The normalized grayscale values increased by as much as 2-fold for all the tested formulations, thus providing a clear indication of the phase separation phenomenon. The time at which the derivative of such grayscale plots reached its maximum was used to determine the time scale of the phase separation process for each composition. The results indicate that this timescale reduces from 3 to 1.75 hours when the PVP fraction is increased from 0 to 10 wt% (Figure 3c). Further additions of PVP beyond 10 wt% lead to a progressive increase of the timescale required for phase separation. Such general trend was observed for all the PR:EG ratios investigated and reveals that there is an optimum PVP fraction to accelerate phase separation in these formulations. On the basis of these turbidity experiments, we identified in a compositional ternary diagram the formulations that undergo faster polymerization-induced phase separation (Figure 3d).

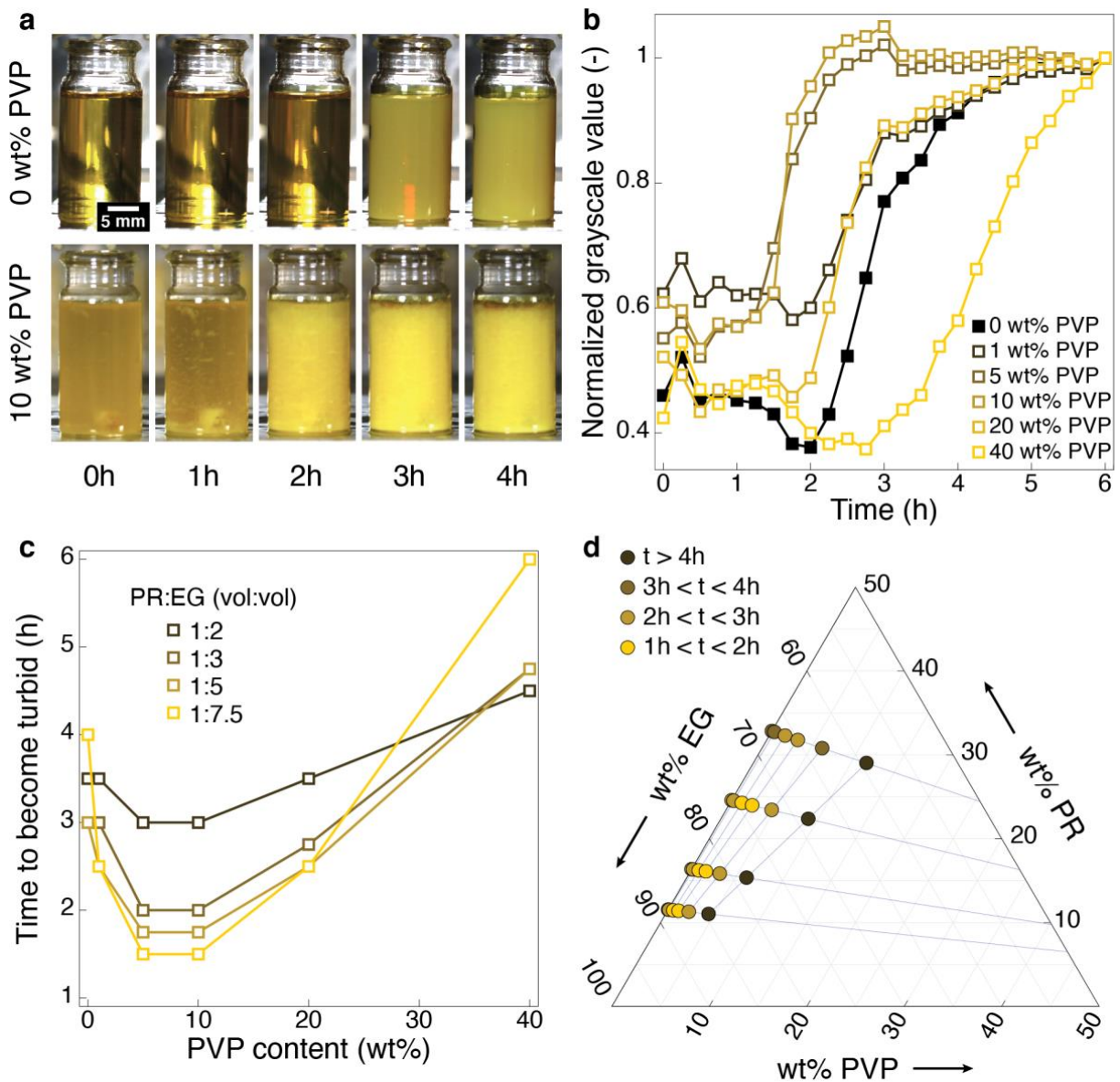

Figure 3: Effect of poly(vinylpyrrolidone) (PVP) on the phase separation of the phenol resin mixture during heating. (a) Time-lapse photographs of mixtures containing 0 and 10 wt% of PVP with respect to phenolic resin for a fixed PR:EG ratio of 1:5. (b) Changes in normalized grayscale value of photographs similar to the one shown in (a) as a function of time for mixtures containing different PVP concentrations and a PR:EG ratio of 1:5. (c) Time for the mixtures to become turbid as a function of their PVP contents. The time to become turbid is determined as the one to reach the maximum of the derivative of the greyscale curves. (d) Compositional ternary diagram indicating the time to become turbid of all the investigated mixtures.

While the role of PVP on the phase separation process is not fully clear, previous research suggests that this polymer binds relatively strongly with polyphenol molecules in other colloidal systems. [43-46] Such attractive interactions have been attributed to hydrophobic forces, complexation, hydrogen bonding, or π-π interactions between the cyclic carbon rings of the two macromolecules. [43, 46] The quick phase separation observed in our PVP-containing mixtures might be induced by these attractive interactions, since these can potentially favor the formation of the phenolic polymer network. In contrast to the fabrication of porous membranes, the phase separation effect investigated here is induced by the polymerization of the phenolic resin and not by the exchange of solvents. This prevents us from drawing conclusions based on the current understanding of the phase separation phenomena involved in the formation of porous polymer membranes. Further research is required to elucidate how PVP affects the mechanism of phase separation during polymerization of the phenolic resin.

The optimum PVP fractions revealed by the phase separation experiments were also found to be very effective in creating a highly porous microstructure after polymerization of the phenolic resin and removal of the ethylene glycol solvent (Figure 4). Scanning electron microscopy images (SEM) of polymerized and dried samples show that the addition of 5–10 wt% PVP results in a high density of micrometer-sized open pores in the phenolic material for all the tested PR:EG ratios. The formation of these open pores is favored in formulations containing higher fractions of ethylene glycol, indicating the importance of this solvent in generating the phase-separated domains that give rise to porosity in the system. Importantly, the size range of the pores generated in compositions containing optimum PVP fractions is clearly much larger than the pores formed within the aerogel phase of ablators reported in the literature. [22, 24, 25] These encouraging experiments set the stage for the development of enhanced carbon ablators through the impregnation of carbon fiber networks with PVP-containing PR-EG mixtures.

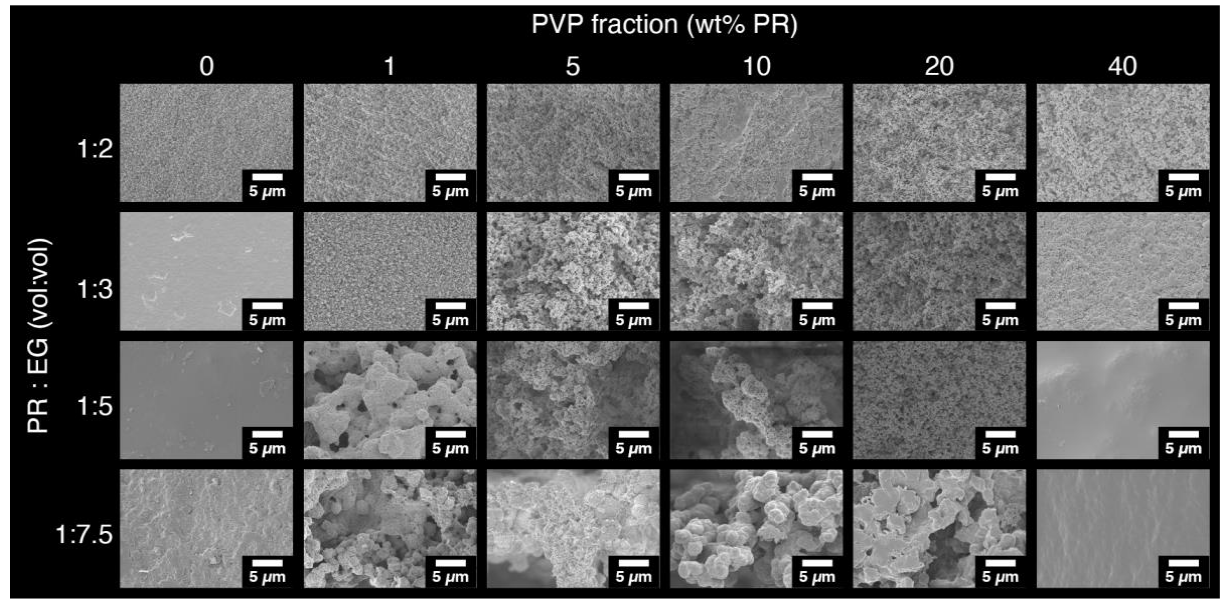

Figure 4: Electron microscopy images of porous samples prepared via the polymerization of resin mixtures containing different fractions of poly(vinylpyrrolidone) (PVP) and relative ratios of phenolic resin (PR) and ethylene glycol (EG).

*3.3. Fabrication of phenolic impregnated carbon ablators*

Phenolic impregnated carbon ablators (PICA) with designed porosity were prepared by infiltrating a carbon fiber network with the solution of phenolic resin, ethylene glycol and poly(vinylpyrrolidone) (Figure 5a). The carbon fiber network is commercially available under the tradename Fiberform. Because of its manufacturing process, it displays carbon fibers oriented within the plane of the monoliths. [47] The carbon fibers are rayon-based and have a diameter of 10-20 $\mu$m. [48, 49] They are bond together by a thin carbonaceous phase generated upon pyrolysis of phenolic resin. [7] The monoliths exhibit a volume fraction of fibers of approximately 10% and an apparent density of 0.18 g/cm$^3$. [49, 50]

For the impregnation process, the pre-assembled carbon fiber monolith is first immersed in the infiltration solution, both of which were pre-heated to 110°C. Based on the earlier experiments, the infiltration solutions contained a PVP fraction corresponding to 10 wt% of the resin and a constant PR:EG volume ratio of 1:5. In addition to providing pore sizes in the desired range, this resin:solvent ratio is particularly suitable because it leads to a low-viscosity solution that can easily be infiltrated in the fiber monolith, while also providing a resin fraction that is sufficiently high to build a mechanically robust porous network between the fibers. The impregnation step was carried out under a reduced pressure of 30 mbar for 30 to 60 minutes to facilitate the infiltration of the resin mixture. The infiltrated mixture was eventually polymerized by increasing the temperature to 150°C and holding it for 12 hours. Finally, the solvent was removed by evaporation at 150°C under vaccum to generate the porous carbon ablator. Porous monoliths prepared with 10 wt% PVP (Figure 5c) could be readily distinguished from other samples (Figures 5b, d) due to their brown-yellow color. Such color is similar to that observed in resins polymerized in the presence of PVP (Figure 3a) and is indicative of the light scattering nature of the monolith.

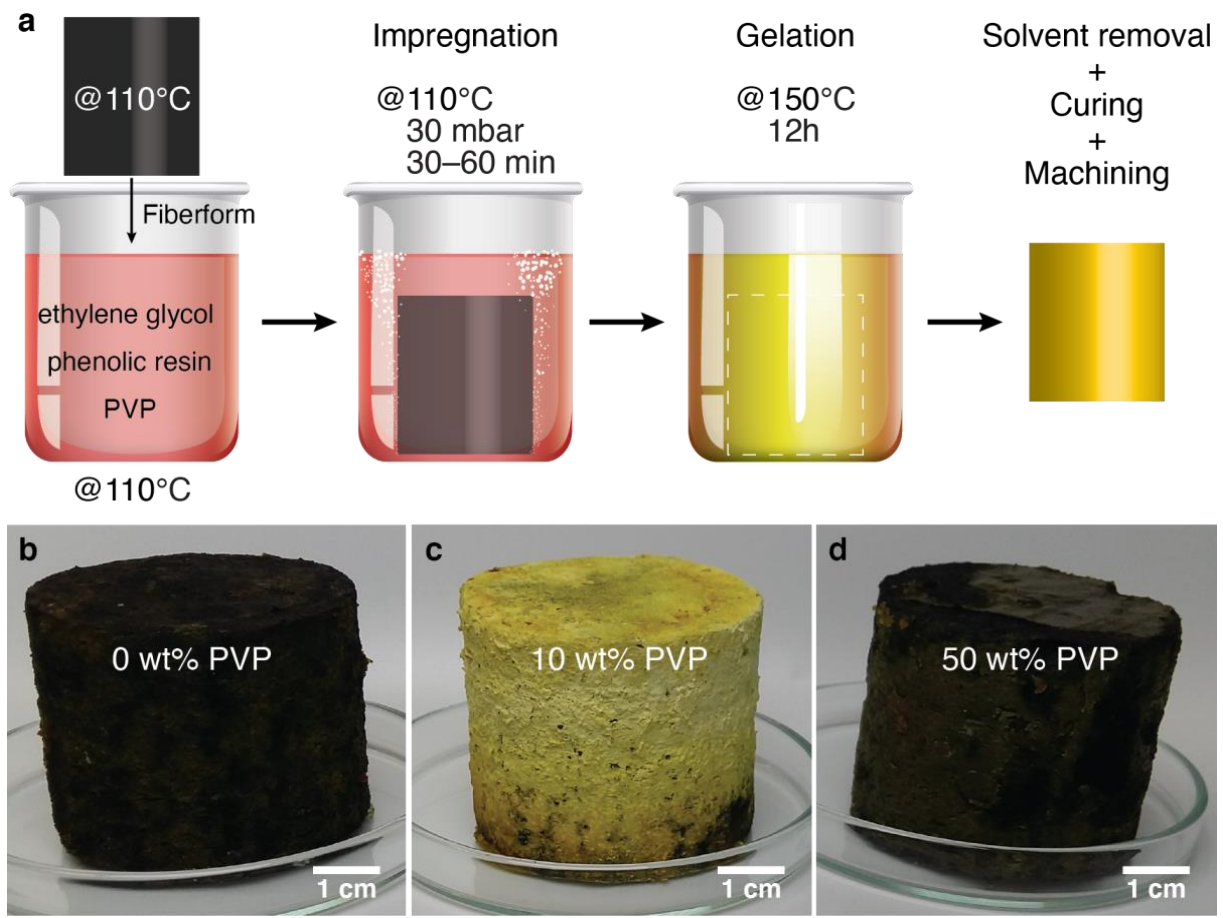

Figure 5: Fabrication of porous carbon ablators. (a) Processing steps used to prepare phenolic-impregnated carbon ablators (PICA) with designed porosity. (b-d) PICA samples obtained after the solvent removal step containing (b) 0, (c) 10 and (d) 50 wt% PVP with respect to the phenolic resin. The yellow color in (c) suggests strong light scattering of the sample and confirms the existence of an optimal PVP content for the creation of pores through phase separation.

### 3.4. *Ablator microstructure and mechanical properties*

The microstructure of the porous ablator is strongly affected by the presence of PVP in the infiltration solution. Optical and electron microscopy images of ablators containing 5–10 wt% PVP reveal that the openings within the carbon fiber network are fully occupied by a space-filling porous phase (Figure 6). For PVP fractions outside this optimum interval, the carbon fibers can be identified more easily in the microscopy images due to the larger openings inside the network. These results agree well with the microstructural analysis of the polymerized resins (Figure 4), thus indicating that the phase separation process observed in the solution also occurs when this solution is infiltrated into the carbon fiber monolith. Such an agreement also suggests that the pore-forming process is sufficiently robust to be applied in different processing scenarios.

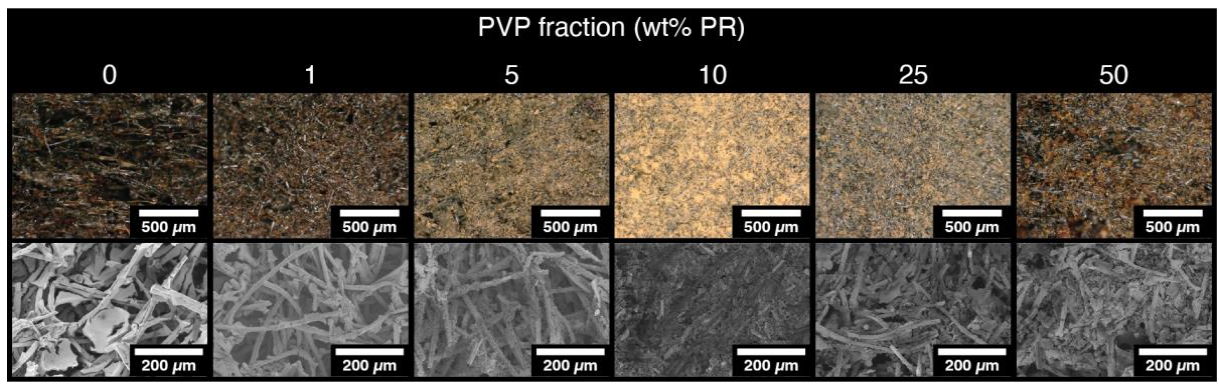

Figure 6: Optical (top row) and electron microscopy (bottom row) images of porous carbon ablators prepared with different PVP fractions and a fixed PR:EG volume ratio of 1:5.

To further characterize the porous structure formed within the carbon fiber network, monoliths with 0 and 10 wt% PVP were also evaluated by BET analysis and mercury intrusion porosimetry (MIP) (Figure 7a,b). BET was used to assess the size distribution and volume of pores smaller than approximately 50 nm, whereas MIP provides information about pores larger than 100 nm. The pore size distribution of the reference PICA sample (0 wt% PVP) displays two main families of pores. Pores with average size of 70 µm are found at larger length scales and correspond to the openings within the carbon fiber network of the ablator. At smaller length scales, the aerogel phase within the fiber network gives rise to pores with size ranging from 5 to 50 nm.

Notably, the addition of 10 wt% PVP into the resin mixture shifts the coarser pores of the reference ablator towards the pore size interval required for enhanced scattering of thermal radiation (0.1–10 µm). Indeed, the MIP analysis shows that the coarse pores found in the PVP-containing specimen lie predominantly within the size range 4–12 µm. In addition to these coarse pores, samples prepared with 10 wt% PVP also feature aerogel-like pores smaller than 10 nm. To estimate the percentage of coarse porosity in the structure, we take the integrals of the porosimetry curves for their whole lengths and for the size interval 0.1–10 µm. The percentage of coarse porosity is defined as the specific volume of pores with diameters in the range 0.1–10 µm divided by the total specific volume of pores. This analysis led to coarse porosity values of 0.7% and 72.3% for specimens with 0 and 10 wt% PVP, respectively. The high percentage of coarse pores observed for the PVP-containing sample indicates that most of the porosity of these structures lies within the size interval that promotes scattering of thermal radiation.

In addition to pore size, the total porosity and mechanical properties of the ablator are crucial to provide effective thermal insulation and sufficient mechanical stability. To evaluate the effect of PVP on the total porosity of the structure, we measured the density and the open porosity of specimens with varying fractions of the polymer additive (Figure 7c). The results show that the density of specimens prepared with up to 25 wt% PVP falls within the range 0.25–0.29 g/cm$^3$, which is close to the reference value of 0.27 g/cm$^3$ measured for the PVP-free reference structure. Higher density variations can be achieved upon changes in the PR:EG volume ratio, here kept constant at 1:5. Mechanical compression tests on samples with different PVP fractions also show that the fracture strength and elastic modulus of the ablator are not affected by the introduction of up to 25 wt% of the polymer additive in the resin formulation (Figure 7d). Overall, our experiments indicate that the incorporation of PVP in the resin reduces the size of coarse pores without significantly affecting the total porosity and the mechanical properties of the ablator.

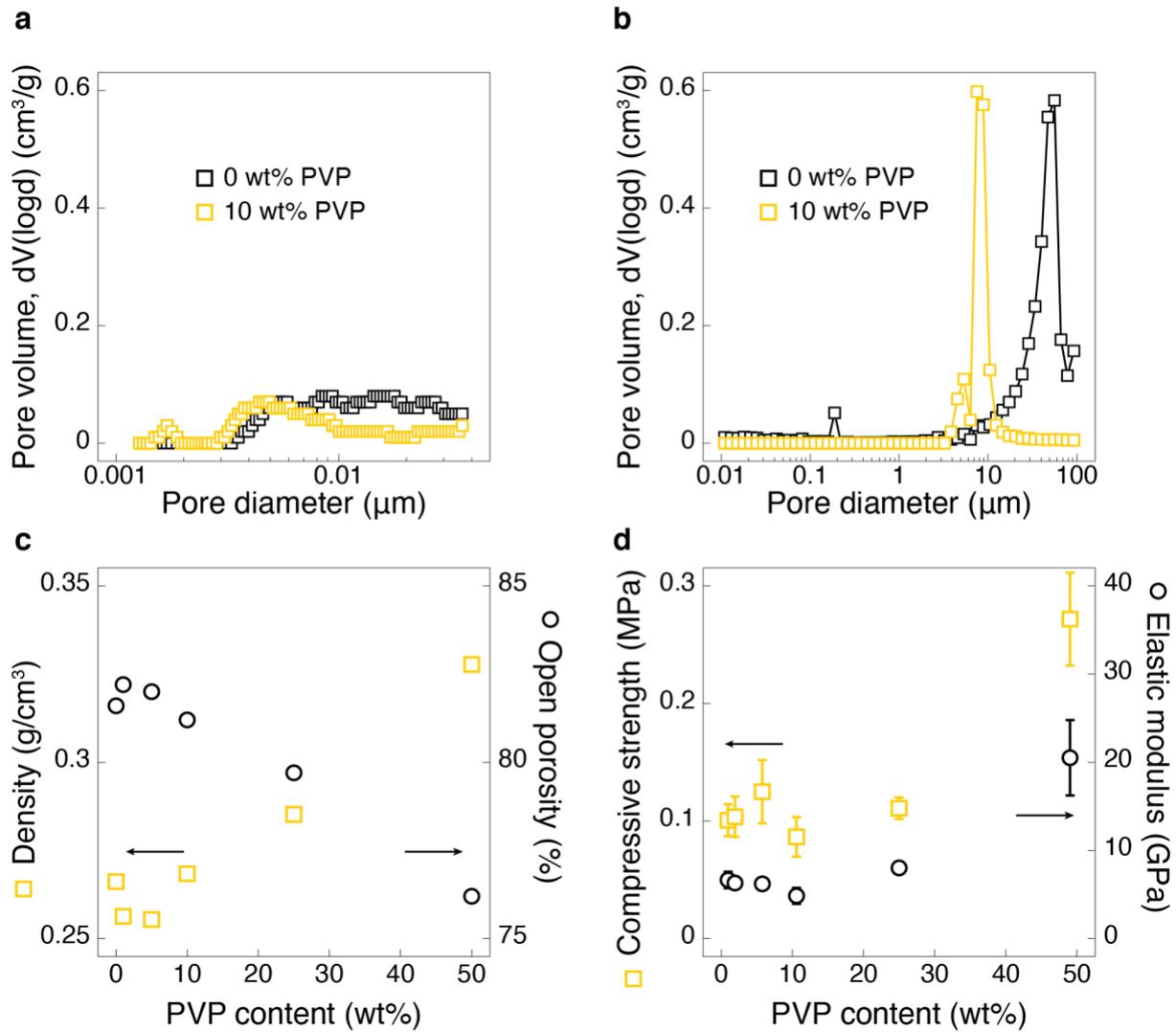

Figure 7: Effect of the PVP fraction on the pore size distribution, porosity and mechanical properties of carbon ablators. Pore size distribution assessed via (a) BET and (b) mercury intrusion porosimetry (MIP) analyses. (c) Density, open porosity, (d) compressive strength and elastic modulus of ablators as a function of the content of PVP present in the initial resin mixture.

### 3.5. *Heat-shielding performance*

Porous structures with pore sizes in the desired scattering range were evaluated in terms of thermal radiation resistance by exposing them to laser radiation at intensities comparable to those expected in the harshest atmospheric re-entry conditions. The measurement setup consists of a laser source, an argon-filled chamber, a thermocouple and digital and infrared cameras (Figure 8a,b). For the test, the sample is positioned on top of a ceramic thermal insulator and exposed to the strong laser beam while the surface temperature is recorded. To quantify the layer of char formed within the structure under radiation conditions comparable to that expected during re-entries in the Earth's atmosphere, [51] we use a laser with wavelength of 1.03 μm.

The tests were performed on specimens with 0 and 10 wt% PVP prepared at PR:EG volume ratios of 1:3 and 1:5. Resin formulations with PR:EG ratios of 1:3 and 1:5 led to sample densities of 0.30 and 0.24±0.01 g/cm$^3$, respectively. Cylindrical specimens with diameter of 43 mm and 31 mm in height were machined directly from impregnated carbon preforms. During the test, samples were irradiated with a laser intensity of 400 W/cm$^2$ for 30 seconds, which is comparable to the conditions often employed for ablation measurements. [22, 25, 52] The distance of the laser source to the sample was adjusted so as to expose a circular area with diameter of 32 mm. Although the customized setup and the testing conditions are different from the arc-jet wind tunnel typically used to simulate re-entry conditions, our experiment is based on previously reported setups [52, 53] and should provide a reasonable indication of the effect of the structure's porosity on the high-temperature performance of the ablator.

Infrared imaging of the sample during exposure to the high-intensity laser beam reveals that the temperature on the surface rises to above the detection limit of 1800°C within less than 1 second for all the tested ablators. A temperature map of the sample surface after 30 seconds of laser exposure indicates that such high temperature was achieved for most of the illuminated area (Figure 8c). Cross-sections of the samples after the laser test allowed us to quantify the amount of polymer that was pyrolyzed into carbon under the laser beam, which is known as the char depth (Figure 8d,e,g,h).

Analysis of the char depth indicates that the presence of 5–20 μm pores in the PVP-containing samples significantly improved the heat-shielding performance of the porous insulator. Indeed, images of the cross-section of laser-tested samples reveal that the addition of PVP reduces by 30–42% the char depth of the porous ablator in comparison to the PVP-free reference material (Figure 8f,i). Such effect occurs for samples with both 0.30 and 0.24±0.01 g/cm$^3$, and was confirmed by independent measurements of the residual fraction of resin along the sample depth (see supporting information, Figure S1). The lower char depths observed for PVP-containing samples likely result from the lower radiative thermal conductivity expected for such porous structures. The assumption underlying this interpretation is that the pore size within the phenolic resin is comparable to that obtained upon pyrolysis of the material at high temperatures. Therefore, our experimental results support the initial hypothesis that the thermal performance of the ablator can be effectively enhanced through the introduction of pores that promote radiation scattering in the insulator at high working temperatures.

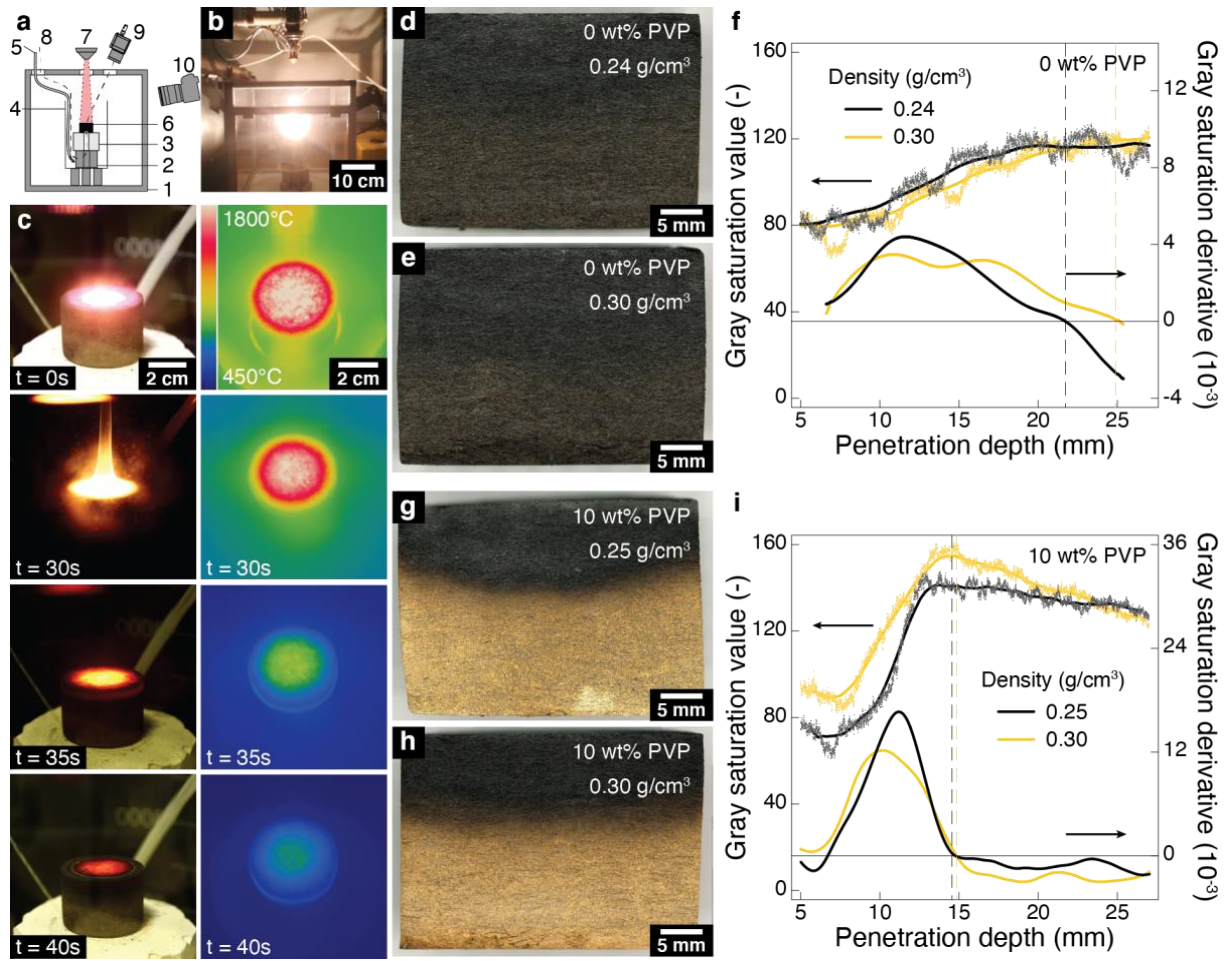

Figure 8: Assessment of the heat-shielding performance of the porous carbon ablators. (a) Schematics of the experimental setup using a laser as radiation source. The setup consists of a (1) test cell, (2) a spacer, (3) a ceramic insulator, (4) a glass beaker, (5) argon tubing, (6) the specimen, (7) a laser nozzle, (8) a K-type thermocouple, (9) an IR camera and (10) a digital camera. (b) Photograph of the setup during a test. (c) Time-lapse photographs of the laser test taken with digital and infrared cameras at 0, 30, 35 and 40 s. (d,e) Cross-sections of laser-tested samples prepared with 0 wt% PVP at densities of (d) 0.24 g/cm$^3$ and (e) 0.30 g/cm$^3$. (f) Gray saturation curves of the images shown in (d) and (e) and their derivatives as a function of penetration depth into the samples. (g,h) Cross-sections of laser-tested samples prepared with 10 wt% PVP at densities of (g) 0.25 g/cm$^3$ and (h) 0.30 g/cm$^3$. (i) Gray saturation curves of the images shown in (g) and (h) and their derivatives as a function of laser penetration depth into the specimens. The results show that penetration depths with 10 wt% PVP (i) are 30–42% lower than the ones with 0 wt% PVP (f).

4. **Conclusions**

Phenolic-impregnated carbon ablators with pores sizes designed to promote radiation scattering display superior heat-shielding performance at high temperatures compared to state-of-the-art thermal protection systems reported in the literature. Analytical models predict that pores ranging from 100 nm to 10 μm should increase the scattering of the radiation expected at 1800 °C and thus decrease the radiative contribution to the thermal conductivity of the porous material at such temperature. We experimentally discovered that the addition of poly(vinylpyrrolidone) (PVP) into a conventional impregnation mixture of phenolic resin and ethylene glycol is an effective approach to create 5–20 μm pores that enhance scattering of thermal radiation. PVP was found to facilitate the polymerization-induced phase separation of the mixture into polymer-rich and solvent-rich domains. This effect might be driven by specific molecular interactions between the PVP macromolecules with the growing phenolic polymer. The phase separation observed in PVP-containing phenolic mixtures can be performed inside carbon fiber networks to produce large-scale thermal insulators of enhanced heat-shielding performance without significantly affecting their porosity and mechanical properties. Laser tests point out that the char depth in insulators prepared with PVP is up to 42% lower compared to that measured in the reference carbon ablator. This improved performance shows that tailoring the pore size of thermal insulators is an attractive strategy to produce effective thermal protection systems while keeping the low cost and material availability of porous carbon ablators reported in the literature. Similar to reflective photonic structures created by living organisms in Nature, the porous architectures responsible for the enhanced thermal performance are generated through a self-assembly process that is easily up-scalable to large structures.

## 5. Materials and Methods

### 5.1. *Materials*

Resole phenolic resin (Cellobond SC1008P™, Hexion), ethylene glycol (>98%, VWR chemicals) and poly(vinylpyrrolidone) ($M_W = 10000$ g/mol, Sigma Aldrich) were used without further purification. A carbon fiber porous monolith (Fiberform®) with an apparent density of 0.18 g/cm³ was supplied by Fiber Materials Incorporated (FMI). Cylindrical specimens with diameter of 50 mm and height of 36 mm were punched directly from carbon fiber monoliths (Fiberform blocks) with the help of a thin-walled steel tube.

### 5.2. *Manufacturing of samples*

Carbon ablators were manufactured by infiltrating the carbon fiber monolith with a solution of phenolic resin dissolved in ethylene glycol with or without poly(vinylpyrrolidone) (PVP). In this process, the carbon fiber monolith was first preheated to 110°C. The phenolic resin and the PVP were dissolved under vigorous stirring in different glass beakers, each one containing half of the ethylene glycol required in the final mixture. These two solutions were brought to 110°C using a silicone oil bath. Afterwards, the PVP solution was carefully poured into the solution containing the phenolic resin. The resulting mixture became fully homogeneous within 10–15 minutes of continuous stirring at 110°C. The preheated carbon fiber monolith was then immersed into the mixture and transferred from the oil bath setup to a vacuum oven at 110°C. Full impregnation of the monolith was facilitated by carefully pulling vacuum until a pressure of 30 mbar was reached. The pressure was kept at 30 mbar for 30–60 minutes until air bubbles could no longer be seen.

To polymerize the phenolic resin, the sample was transferred from the vacuum oven to a forced convection oven preheated to 150°C, where it was kept for 24h. The sample turned yellow upon polymerization. Subsequently, the sample was allowed to cool down to 50°C inside the oven. Next, the sample was extracted from the glass beaker and the excess of polymerized resin around the impregnated ablator was removed to allow the ethylene glycol to evaporate more easily.

For the evaporation step, the ablator was placed into a vacuum oven at 150°C and a pressure of 50 mbar for 24h. The dried ablator was then placed into a forced convection oven for 12h at 180°C for the final curing step. The outermost layer of cured resin around the ablator was removed by grinding the specimen with a 320-grit SiC paper until a final diameter and height of 43 and 31 mm were, respectively, reached (Figure 5a).

*5.3.Characterization*

The structure of polymerized resin samples and carbon ablators was characterized using optical microscopy (Keyence VHX-6000 digital microscope) and scanning electron microscopy (SEM, LEO 1530 Gemini microscope). The SEM was equipped with an in-lens detector and was operated at an acceleration voltage of 2 kV, an aperture size of 30 μm and a working distance of 3–5 mm (Figure 6).

Nitrogen sorption curves were measured at -196°C with a gas sorption analyzer (Quantachrome autosorb iQ) after outgassing the samples under vacuum for 18 h at 60°C. The surface area was evaluated by the Brunauer-Emmett-Teller (BET) method and the pore size distribution was determined by a non-local Density Functional Theory (NLDFT) calculation model for nitrogen at -196°C on cylindrical pores in carbon (Figure 7a). The mercury intrusion porosimetry analysis was carried out by pressing non-wetting liquid mercury into the sample (Pascal 140-240/440 EVO, Thermo Fisher Scientific). The required intrusion pressure was used to calculate the diameter of the filled pores (Figure 7b).

The density and the open porosity of the carbon ablators were calculated using the Archimedes' method. The dry, wet and immersed masses of the samples were measured in ethanol at 22°C using a density measurement kit (Mettler Toledo) mounted on a precision balance (Figure 7c).

To perform compression tests, cylindrical specimens were punched from the ablator using a steel tube with 9.5 mm diameter. The final diameter and height of the specimens varied from 8 to 9 mm and from 9 to 17 mm, respectively. Compression tests were carried out on a mechanical universal tester (Shimadzu AGS-X) equipped with two steel plates and a 1 kN load cell. The compression speed was fixed at 2 mm/min. The elastic modulus and the compressive strength of the specimens were directly calculated from, respectively, the slope and the maximum stress value of the stress-strain curves within the linear elastic region (Figure 7d).

For the laser tests, a 5-L glass beaker was placed inside a 50-L cubic cell made with a metallic frame and PMMA walls. The sample was placed on a ceramic block to protect the cell from thermal damage. The tests were conducted with a disk laser (Trumpf® TruDISK 4001, maximum $P = 3\ kW$, $\lambda = 1.03$ μm) equipped with a 400 μm fiber optics cable and a nozzle (Oerlikon Metco) positioned directly above the sample (Figure 8a,b). To obtain an average heat flux of 400 W/cm$^2$ for 30 seconds at maximum laser power, the working distance between the laser and the sample was set to approximately 345 mm. At this distance, the pilot laser used as a reference had a diameter of 32 mm on the specimen's surface. The top surface temperature was measured using an infrared camera (Optris PI 1M) with the surface emissivity set to 0.9 (Figure 8c). [7] Each test started by flushing the test cell with argon at a rate of 12 L/min for 5 min to create an inert atmosphere. During the test, the laser nozzle was protected by an argon flow of 12 L/min and the argon flow to the test cell was reduced to 3 L/min. The ventilation system of the laser room was kept off the whole time to ensure the formation of a sufficient argon blanket. The samples were irradiated with 3 kW for 30 seconds, resulting in an overall radiation energy of 11.92 kJ/cm$^2$.

The char depth in the laser-tested specimens was defined as the distance from the test surface at which an alteration of the optical or structural appearance of the sample was visible. Importantly, no ablation occurs during the laser tests, since the samples are not exposed to a high-velocity gas flux during the measurements. The char depth was obtained by image analysis of stitched optical microscope images (50 X magnification ) (Figure 8d,e,g,h). For a detailed analysis, the images were converted to the HSB (Hue, Saturation, Brightness) color space using the software ImageJ. [54] The vertical gray value profile of the saturation channel, referred to here as the gray saturation profile, was plotted for a square of 30 mm starting at the upper surface of the sample's cross section. This allowed us to average the gray values horizontally. The char depth was taken as the distance at which the derivative of smoothened gray saturation curves approaches zero (Figure 8f,i).

To confirm the results obtained by image analysis, we also estimated the char depth from thermogravimetric analysis (TGA). To this end, samples with a density of 0.30 g/cm$^3$ were cut in 1 mm-thick slices along their depths and a TGA measurement of each slice was performed. The slices were heated in air from 25°C to 850°C at heating and cooling rates of 10°C/min (NETZSCH STA 449 C and Mettler Toledo TGA/DSC 3+ STARe System). The TGA data was used to estimate the mass fraction of phenolic resin in each slice, given by the mass loss associated with the thermal decomposition of the resin divided by its initial mass. The char depth of the sample was taken as the point at which the mass fraction of the phenolic resin becomes constant and its derivative with respect to depth approaches zero (Figure S1).

**Acknowledgements**

This work was financed by the Swiss National Science Foundation within the framework of the project 200021_160184. The authors wish to acknowledge Florian Wirth, Dominik Keller and Joseph Stirnimann (Inspire AG, ETH Zürich) for the technical support with the laser tests, and Laura Conti and Thomas Graule (Laboratory for High Performance Ceramics, EMPA) for the mercury intrusion porosimetry analyses. We are also grateful to Julia Manser and Amine Layadi (ETH Zürich) for contributing to the processing of the composites, and to Andrey V. Gusarov (Moscow State University of Technology) for the advice given to set up the laser testing conditions.

**Supporting Information**

**Carbon ablators with porosity designed for enhanced aerospace thermal protection**

Erik Poloni[1], Florian Bouville[1+], Alexander L. Schmid[1], Pedro I. B. G. B. Pelissari[2], Victor C. Pandolfelli[2], Marcelo L. C. Sousa[1], Elena Tervoort[1], George Christidis[3], Valery Shklover[3], Juerg Leuthold[3], André R. Studart[1]

[1] Complex Materials, Department of Materials, ETH Zürich, 8093 Zürich, Switzerland

[2] Federal University of São Carlos, Materials Engineering Department, São Carlos, SP, Brazil

[3] Institute of Electromagnetic Fields, ETH Zürich, 8092 Zurich, Switzerland

[+] Now at: Centre for Advanced Structural Ceramics, Department of Materials, Imperial College London, United Kingdom

*Relation between the Rosseland extinction coefficient ($\beta_R$) and the spectral extinction coefficient ($\beta_{ext}$)*

The Rosseland extinction coefficient is expressed by: [32]

$$\frac{1}{\beta_R} = \frac{\int \beta_{ext}^{-1} \frac{dB(T)}{dT} d\lambda}{\int \frac{dB(T)}{dT} d\lambda}$$

where $\beta_{ext}$ is the spectral extinction coefficient and $\frac{dB(T)}{dT}$ is the derivative of Planck's thermal radiation spectrum, $B(T)$. For cases in which radiation scattering dominates over absorption phenomena, one obtains: $\beta_{ext} = Q_s(\lambda, r) N_p A_p$, where $Q_s(\lambda, r)$ is the scattering efficiency for a pore size $r$ and wavelength $\lambda$, $N_p$ is the number of pores per unit volume of material and $A_p$ is the cross-sectional area of the pore.

Planck's thermal radiation spectrum, $B(T)$, is expressed by:

$$B(T) = \frac{2hc^2}{\lambda^5} \frac{1}{e^{\frac{hc}{K_B T \lambda}} - 1}$$

where $h$ is Planck's constant, $c$ is the light speed in vacuum and $K_B$ is Boltzmann's constant.

The estimation of the Rosseland coefficient ($\beta_R$) using the equations above can be simplified if one assumes a fixed wavelength corresponding to the maximal thermal radiation emitted by a black body at a certain temperature. This assumption yields:

$$\beta_R = \beta_{ext}$$

*Absorption and scattering contributions to the thermal conductivity*

To compare the contributions of absorption and scattering to the radiative thermal conductivity of the ablator, we compare the absorption coefficient ($\beta_a$) for carbon reported in the literature [33] with the scattering coefficient ($\beta_s$) estimated in our work (Figure 1b).

In a previous study, experiments have shown that the mass specific absorption coefficient of carbon aerogels is approximately 150–200 m²/kg for wavelengths in the range 2–16 $\mu$m at room temperature. [33] Taking the specific gravity of 2 g/cm³ for carbon, these values correspond to $\beta_a$ ranging from 0.3 to 0.4 $\mu m^{-1}$.

The theoretical analysis of the scattering coefficient of carbon ablators with pore sizes in the optimum range (0.1–10 μm) at an arbitrary temperature of 1800°C (Figure 1b) predicts that the $\beta_s$ value increases from 0.35 $\mu m^{-1}$ at $2r = 0.1 \mu m$ to a maximum of 3.7 $\mu m^{-1}$ at $2r = 0.55 \mu m$, before it decreases again to 0.25 $\mu m^{-1}$ at $2r = 10$ $\mu m$. Although data for carbon absorption was obtained at room temperature, this comparison suggests that the contribution of scattering in carbon ablators with pore sizes within the optimum range is on the same order of magnitude or 10-fold higher than the contribution expected from absorption. Such comparison reveals that our model assumption that scattering dominates the radiative contribution $k_{rad}$ will lead to more accurate thermal conductivity predictions for pore diameters around 0.4 μm and less precise estimations at the boundary pore diameters of 0.1 μm and 10 μm. Despite the poor predictability at the boundary pore diameters, our assumption is still justified by the scarcity of experimental data for the absorption of carbon at the wavelengths and temperatures relevant for this work, and by the lack of simple models that take into account the effect of absorption of the continuous phase on the scattering behavior of porous materials.

*Calculation of the conductive contribution ($k_{cond}$) to the thermal conductivity*

The contribution of solid and gas conduction to the thermal conductivity of a porous material ($k_{cond}$) is given by: [32]

$$k_{cond} = k_s M (1 - P)^{3/2} + k_g P^{1/4}$$

where $P$ is the porosity of the material, $M$ is a constant associated with grain boundaries and microcracks, $k_s$ is the thermal conductivity of the solid phase and $k_g$ is the thermal conductivity of the gas phase. The term $k_g$ is calculated by

$$k_g = \frac{k_{atm}}{1 + C\left(\frac{T}{P_g d}\right)}$$

where $k_{atm}$ is the gas thermal conductivity in normal atmospheric conditions, $C$ is a constant related to the gas chemical composition, $T$ is the absolute temperature, $P_g$ is the gas pressure and $d$ is the pore diameter.

The $k_{cond}$ values estimated in this study (Figure 1c,d) were calculated assuming $P = 0.8$, $k_s M = 1.397 \frac{W}{mK}$, [7] $k_{atm} = 0.016 \frac{W}{mK}$, $C = 0.000025$ and $P_g = 101325\ Pa$.

*Confirmation of heat-shielding performance of ablators by TGA analysis*

In addition to image analysis, thermogravimetric measurements were also used to estimate the char depth in samples subjected to the laser tests (Figure 8). The TGA data confirms that the char depth in ablators prepared with 10 wt% PVP is significantly lower than that of the reference specimen without PVP (Figure S1). In this analysis, slices of 1 mm were cut from samples with a density of 0.30 g/cm³ and were characterized via TGA. Slices taken deeper into the samples show higher mass losses at 600 °C (Figure S1a,b). This means that the resin fraction in the ablator increases with penetration depth. The char depth was estimated as the depth at which the derivative of the resin fraction curve approaches zero (Figure S1c).

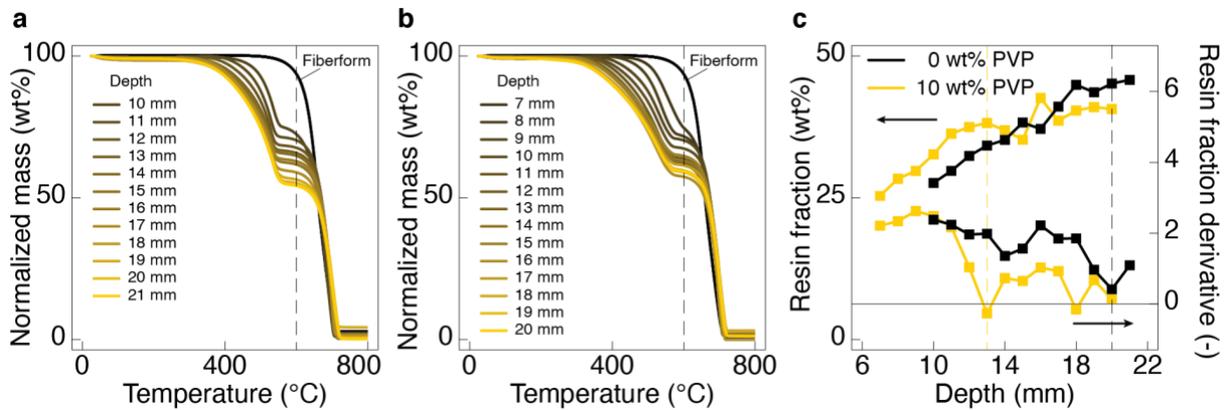

Figure S1: Normalized mass as a function of temperature obtained by TGA of slices of laser-tested ablators with density of 0.30 g/cm³ (a) without PVP and (b) with 10 wt% PVP. (c) Mass fraction of phenolic resin as a function of sample depth for ablators with and without PVP. The char depth is defined as the depth at which the resin fraction is constant.